\documentclass[12pt]{article}
\usepackage{epsfig,epsf}
\def\beq{\begin{equation}}
\def\eeq{\end{equation}}
\def\bea{\begin{eqnarray}}
\def\eea{\end{eqnarray}}
\def\eq#1{{Eq.~(\ref{#1})}}
\def\fig#1{{Fig.~\ref{#1}}}
\newcommand{\bas}{\bar{\alpha}_S}
\newcommand{\as}{\alpha_S}

\newcommand{\pz}{{\phi'_z}}
\newcommand{\pzn}{{\phi'_{z,0}}}

\parskip=\itemsep               

\newcommand{\beqar}[1]{\begin{eqnarray}\label{#1}}
\newcommand{\eeqar}{\end{eqnarray}}
\newcommand{\h}{\frac{1}{2}}

\newcommand{\z}{\hat{z}}
\newcommand{\N}{\hat{N}}
\renewcommand{\L}{\left(}
\newcommand{\R}{\right)}
\renewcommand{\theequation}{\thesection.\arabic{equation}}

\begin{document}
\def\thefootnote{\fnsymbol{footnote}} 
\makeatletter
\def\eqnumsection{\@addtoreset{equation}{section}
\def\theequation{\arabic{section}.\arabic{equation}}}

\def\appendixes{\par\setcounter{section}{0}
\setcounter{subsection}{0}\setcounter{equation}{0}
\@addtoreset{equation}{section}
\def\thesection{A\arabic{section}}
\def\theequation{A\arabic{section}.\arabic{equation}}}

\makeatother

%
%
%
\noindent
\begin{flushright}
\parbox[t]{10em}{ \tt TAUP \,\,2803/2004\\
 \today }
\end{flushright}
\vspace{1cm}
\begin{center}
{\LARGE  \bf Solution  to the Balitsky-Kovchegov equation}\\
~ \\
{\LARGE  \bf  in the saturation domain}
\\

\vskip1cm {\large \bf ~M. ~Kozlov $ {}^{\dagger}$ \footnotetext{${}^{\ddagger}$ \,\,Email:
kozlov@post.tau.ac.il} and ~E. ~Levin ${}^{ \star}$ \footnotetext{${}^{\,\star}$ \,\,Email:
leving@post.tau.ac.il, levin@mail.desy.de }} \vskip1cm

{\it  HEP Department}\\
{\it School of Physics and Astronomy}\\
{\it Raymond and Beverly Sackler Faculty of Exact Science}\\
{\it Tel Aviv University, Tel Aviv, 69978, Israel}\\
\vskip0.3cm

\end{center}  
\bigskip
\begin{abstract} 	
The solution to the Balitsky-Kovchegov equation is found in the deep saturation domain.
The controversy between different  approaches regarding  the asymptotic behaviour of  the scattering 
amplitude is solved. It is shown that the dipole amplitude behaves as $ 1 - \exp \left( - z + \ln z 
\right)$ with $ z = \ln (r^2 Q^2_s)$ ($ r$ -size of the dipole, $Q_s$ is the saturation scale)  in 
the deep saturation region. This solution is developed  from the scaling solution to the homogeneous 
Balitsky-Kovchegov equation.
The  dangers associated with making   simplifications in the 
BFKL kernel, to investigate  the asymptotic behaviour of the scattering amplitude, is pointed out  . 
In particular, the 
fact that the  Balitsky-Kovchegov equation belongs to the  Fisher-Kolmogorov-Petrovsky-Piscounov 
-type of equation, needs  further careful investigation.

\end{abstract}

\newpage 
\eqnumsection
\def\thefootnote{\arabic{footnote}} 
\section{Introduction}

\label{sec:Introduction}
The main objective of this paper is to find the solution to the non-linear Balitsky-Kovchegov equation 
\cite{B,K} in the saturation domain. There exist two solutions for dipole scattering amplitude $N$ 
which were found in Refs. 
\cite{LT,BKL,KL}. They can be presented 
in the form:
\beq \label{SOL}
N\L Y,r;b \R\,\,=\,\,1\,\,-\,\,e^{- \phi(z)}
\eeq
where 
\beq \label{Z}
z\,\,\,=\,\,\ln(Q^2_s(Y;b)\,r^2)\,\,=\,\,\as\,C\,Y\,\,+\,\,\ln\L r^2/r^2_0(b)\R 
\eeq
and  $Q_s$ is a saturation scale \cite{GLR,MUQI,MV,BALE}.
Constant $C$ in \eq{Z} is defined as
\beq \label{C}
C\,\,=\,\,\frac{\chi\L \gamma_{cr}\R}{ 1 -  \gamma_{cr}}
\eeq
 and $\chi$ is the BFKL kernel \cite{BFKL}
\beq \label{CHI}
\chi\L \gamma \R \,\,=\,\,2\,\psi(1) \,\,-\,\,\psi(\gamma) \,\,-\,\,\psi(1 - \gamma)
\eeq
where $\psi(\gamma)\,=\,d \ln \Gamma(\gamma)/d \gamma$
and $\Gamma(\gamma)$ is  the Euler gamma function.

The value for the critical anomalous dimension $\gamma_{cr}$  is determined by the equation 
\cite{GLR,BALE,MUT}
\beq \label{GAMMACR}
\frac{\chi\L \gamma_{cr}\R}{ 1 -  \gamma_{cr}}\,\,=\,\,-\,\frac{d \chi\L \gamma_{cr}\R}{d 
\,\gamma_{cr}}\,.
\eeq

In the first solution \cite{LT}
\beq \label{SOL1}
\phi(z)\,\,\,=\,\,\,\frac{z^2}{2\,C}
\eeq
while the second one has the form \cite{BKL,KL}
\beq \label{SOL2}
\phi(z)\,\,\,=\,\,\,z \,\,-\,\,\ln(z)\,\,.
\eeq
These two solutions lead to quite different approaches to the saturation boundary $N 
\,\,\rightarrow\,\,1$. However, since the equation is a non-linear one,  we cannot claim that only 
\eq{SOL2} survives at high energies.

\eq{SOL1} is well accepted by the experts,  while \eq{SOL2} is still considered by the experts as  
shaky, 
 mostly because its  derivation has not reached  a  stage of   transparency as the first 
solution.

In this paper we present (i) the derivation of both solutions in the framework of the same method; (ii)   
a simple explanation of both solutions; and (iii) the general form of 
approaching the unitarity limit.

\section{ Solution to the Balitsky-Kovchegov equation (general approach)}

\subsection{Equation}
The Balitsky-Kovchegov equation\cite{B,K}  which we  solve in this section, has a form:
\beq \label{BK}
\frac{\partial N\L r,Y;b \R}{\partial\,Y}\,
\,\,
=\,\,\frac{C_F\,\as}{\pi^2}\,\,\int\,\frac{d^2 r'\,r^2}{(\vec{r}\,-\,\vec{r}')^2\,r'^2}\,
\eeq
$$
\L \,\,2  N\L r',Y;\vec{b} + \h\,(\vec{r} - \vec{r}\;')\R
\,\,-\,\,N\L r,Y;b \R\,\,-\,\,  N\L r',Y;\vec{b} 
- \h\,(\vec{r} - \vec{r}\;') \R\, N\L  
 \vec{r} -
\vec{r}\;',Y;\vec{b} - \h \vec{r}\;' \R  \,
\R
$$

where $N\L r,Y;b \R$ is the  scattering amplitude of interaction for the dipole with the size $r$ 
and rapidity $Y = \ln(1/x)$ ($x$ is the Bjorken variable),  at impact parameter $b$.

  It is  useful to consider the non-linear equation in a mixed representation, fixing the  
impact parameter $b$,  and  introducing the transverse momenta as conjugate variable to the  dipole 
sizes. 
The relations  between these two representations are given by the following equations 
\bea 
N(r,y;b)\,\,\,&=&\,\,\,\,r^2\,\int^{\infty}_0
 \,k d k \,J_0(k\,r)\,\,\tilde{N}(k,y;b)\,\,;\label{MR1}\\
\tilde{N}(k,y;b)\,\,\,&=&\,\,\,\,\int^{\infty}_0\,\frac{d 
r}{r}\,\,J_0(k\,r)\,\,N(r,y;b)\,\,; \label{MR2}
\eea

 In this representation the non-linear equation reduces to the form \cite{GLR,MP,BKL}
\beq \label{BKMR}
\frac{\partial\,\tilde{N}(k,y;b)}{\partial 
y}\,\,\,\,=\,\,\,\,\bas\,\left(\,\chi(\hat{\gamma}(\xi))\,\tilde{N}(k,y;b)\,\,\,-
\,\,\,\tilde{N}^2(k,y;b)\,\right)
\eeq
and  $\chi(\hat{\gamma}(\xi))$ is an operator defined as

\beq \label{G}
\hat{\gamma}(\xi)\,\,\,=\,\,1\,\,\,+\,\,\,\frac{\partial}{\partial\,\xi}
\eeq

where $\xi \,\,=\,\,\,\ln(k^2\,R^2)$,  and   $k$ is  the  conjugate variable to the 
colour  dipole 
size and $R$ is the size of the target.  In this definition of the variable $\xi$ we implicitly 
assume that $b \,\ll \,R$, and the amplitude $N$ does not depend on $b$. The alternative approach for 
large values of the impact parameter was developed in Ref. \cite{BKL}, where the definition of 
variable $\xi$ is quite different.

\subsection{Goal and assumptions}
Our main goal is to find the solution in the saturation region where $r^2\,Q^2_s\,\,\gg\,\,1$. 
Due to $s$-channel unitarity constraint \cite{FROI} \footnote{See also Ref. \cite{FRDIS} for an 
application 
of 
the Froissart boundary for hard processes} the scattering amplitude in space representation ($N$) 
should be less than 1. In the momentum representation this limit means that 
$\tilde{N}\,\,\rightarrow\,\,\,\h\,\xi$. We can obtain this estimate just using \eq{MR2}, and 
noticing that for $r\,>\,1/k$ the integral over $r$ is small due to oscillating behaviour of $J_0$ 
in \eq{MR2}.  Therefore, for small $k$  \eq{MR2} reduces to
\beq \label{ASYMM}
\tilde{N}(k,y;b)\,\,\,=\,\,\,\int^{\frac{1}{k}}_0\,\frac{d 
r}{r}\,N(r,y;b)\,\,=\,\,\h\,\int^{\infty}_\xi\,d 
\ln 
(1/r^2) \,N(r,y;b)
\eeq

Assuming that $N$ =1 for all $r\,>\,1/Q_s$ ($ \ln(1/r)\,<\,\ln( Q_s)$)   we obtain from \eq{ASYMM}
that
\beq \label{ASM}
\tilde{N}(k,y;b)\,\,\,\rightarrow\,\,\,\h\,\ln\L Q^2_s/k^2 \R
\eeq

Therefore we know the asymptotic behaviour of the amplitude, and we need to find how our function 
approaches \eq{ASM}. 

In the saturation region we
 expect the so called   geometrical scaling behaviour of the scattering amplitude which was proven 
in Ref. \cite{BALE}, for such  equations (see also Ref. \cite{MP} for more general arguments 
and a  more rigorous  proof, and Ref. \cite{IIM} for an observation that geometrical scaling behaviour 
could be correct, even in the part of perturbative QCD kinematic  region).
 It means that $\tilde{N}(k,y;b)$ is a function of the single variable 
\beq \label{ZMR}
\z\,\,\,=\,\,\ln\L Q^2_s(y,b)/k^2 \R \,\,=\,\,\,\bas \,\frac{\chi(\gamma_{cr})}{1 
\,-\,\gamma_{cr}}\,\,(\,Y\,\,-\,\,Y_0\,)\,\,\,-\,\,\xi\,\,-\,\,\beta(b)\,\,;
\eeq  
where $\gamma_{cr}$  is a solution of  \eq{GAMMACR}  \cite{GLR,MUT,MP} and $Y_0$ is the initial 
rapidity. Function $\beta(b)$ depends on impact parameter, but we will  not discuss it  here.

 Introducing a function $\phi(z)$  we are looking for the solution of the equation in the form 
\beq \label{PHIMR}
\tilde{N}(\z)\,\,\,=\,\,\,\h \,\int^{\z}\,\,d\,z'\,\,\left(\,1\,\,\,-\,\,\,e^{ - 
\,\phi(z')}\,\right)\,\,;
\eeq
\eq{PHIMR} includes the geometrical scaling behaviour and leads to the asymptotic behaviour of \eq{ASM}.

Our assumption that
 function $\phi$ is a smooth function, such that $\phi_{z z} 
\,\,\ll\,\,\phi_z\,\phi_z$ where we denote $\phi_z = d \phi/d z$ and $\phi_{z z} = d^2 \phi/(d z)^2$ 
is essential..
This property allows us to rewrite 
\beq \label{SCDER}
\frac{d^n}{(d z)^n}\,e^{- \phi(z)}\,\,\,=\,\,\,( - \phi_z)^n \,e^{- \phi(z)}
\eeq

\eq{SCDER} means that we  can use the semi-classical approach for the solution to  \eq{BK} \cite{BKL}.

\subsection{Reduction of  the Balitsky-Kovchegov equation to the equation in one variable in the 
saturation 
domain}

Substituting in  \eq{BKMR}  $\tilde{N}$ in the form of \eq{PHIMR},  and replacing $Y$ by $\hat{z}$ 
we obtain 
\beq \label{NLEZ}
\bas \,\frac{\chi(\gamma_{cr})}{1
\,-\,\gamma_{cr}}\,\,\frac{d\tilde{N}(\z)}{d \z }\,\,=\,\,
 \bas 
\left(
\,\,\chi(1 -f)\,\,\tilde{N}(\z)\,\,-\,\,\tilde{N}^2(\z)\,\,\right)
\eeq
where  
$ f$ denotes $f = d/d \z = -\,\partial/\partial \xi$ in \eq{NLEZ}.

Differentiating both part of \eq{NLEZ}with respect to $z$ we reduce \eq{NLEZ} to the form
\bea \label{NLEZ1}
\- \h e^{ - \phi(\hat{z})}\,\,\pz(\hat{z})\,\,&=&\,\, \,f\chi(1 -f)\,\N(\z)
\,\,\,-\,\,\N(\z) \L 1\,\,-\,\,e^{-\phi(z)} \R \nonumber \\
 &=&\,\,\h\,(f\,\chi(1 -f)\,\,-\,\,1)\,\N(\z)\,\,+\,\,\N(\z)\,e^{-\phi(z)}
\eea

 An important property of function 
$f\,\chi(1 -f)\,\,-\,\,1$,  is the fact that at 
small $f$  it has an expansion that starts 
\footnote{It should be stressed that the 
simplified 
model for $\chi(1 -f) = \frac{1}{f (1 - f)}$ does not have this property. This is an explanation why in 
Ref. \cite{LT} where this model was used, the solution was missed  as  we will discuss below.}  from 
$f^3$. Since in our case $f$ is operator   $f \equiv \frac{d}{d z}$, it means that the operator  
$f\,\chi(1 
-f)\,\,-\,\,1$ contains the third and higher derivatives with respect to $z$. Therefore,
  one can see that the first term on 
r.h.s. of \eq{NLEZ1} is proportional to $e^{ -\phi(\z) }$ ( see \eq{SCDER}).  Canceling 
$\,e^{ -\phi}$ on both sides 
of 
\eq{NLEZ1},  and  once more taking the  derivative with respect to $\hat{z}$ we reduce \eq{NLEZ1}
to the form:
\bea 
\frac{\chi(\gamma_{cr})}{1\,-\,\gamma_{cr}}\,\,\frac{d^2\,\phi}{(d \z)^2}
\,\,\,&=&\,\,\,\left(\,1\,\,-\,\,e^{- \phi(\z)}\,\right)\,\,-\,\,\frac{d \,L(\phi_z)}{d 
\,\phi_z}\,\frac{d^2\,\phi}{(d \z)^2}\,\,;\label{NLEPHI1}\\
L(\phi_z)\,\,\,&=&\,\,\,\frac{\phi_z\,\,\chi\left(1\,\,-\,\,\phi_z 
\right)\,\,\,-\,\,\,1}{\phi_z}\,\,;\label{NLEPHI2}
\eea

Function $d L /d \pz$ decreases  at large values of the argument but has double pole singularities 
in all integer points ($\pz = 1,2,3, \dots$ ).

\subsection{Two solutions.}

The existence of two solutions with sufficiently different forms  of the dipole amplitude approaching  
its 
asymptotic value ($N = 1$ in space representation and $\tilde{N} = \h \ln (Q^2_s/k^2)$ in the 
momentum representation) can be seen directly from \eq{NLEPHI1}. Indeed, we expect that 
$\phi(z)$  is large at large values of $z$ and , therefore, we can neglect the term $e^{-\phi(\z)}$ 
in  \eq{NLEPHI1}. The first solution can be obtained from  \eq{NLEPHI1} assuming that $dL/d \pz$ 
gives a small contribution while $\phi'(z)$ is large.
 If it is so  \eq{NLEPHI1} reduces to the simple equation
\beq \label{EQSL1}
\frac{\chi(\gamma_{cr})}{1\,-\,\gamma_{cr}}\,\,\frac{d^2\,\phi}{(d \z)^2}\,\,=\,\,1
\eeq
which leads to  
\beq \label{EQSL2}
\phi(\hat{z})\,\,=\,\,\frac{1\,-\,\gamma_{cr}}{\chi(\gamma_{cr})}\,\,\frac{\z^2}{2}\,\,
\eeq
at large $\z$.
The clearest case when we, indeed have this solution, is the simplification of the BFKL kernel which 
was considered in Ref. \cite{LT}. Namely, the kernel was taken as
\beq \label{SMK}
\omega(\gamma)\,=\,\frac{\as \,N_c}{\pi}\,\left\{\begin{array}{c}
\,\,\,\,\,\,\frac{1}{\gamma}\,\,\hspace*{1cm}\mbox{for}\,\,r^2\,Q^2_s\,<\,1\,; \\ \\ \\
\,\frac{1}{1\,-\,\gamma}\,\,\hspace*{1cm} \mbox{for}\,\,r^2\,Q^2_s\,>\,1\,;
\end{array} \right.
\eeq
 instead of the full BFKL kernel $\omega(\gamma)\,=\,\frac{\as \,N_c}{\pi}\,\,\chi(\gamma)$.
  
In this model function $L$ is equal to zero and \eq{EQSL2} gives the only solution in the saturation 
domain.

For the full BFKL kernel there exists another possibility to have a solution: the term with $dL/d 
\pz$ compensates 1 in the r.h.s. of \eq{NLEPHI1},  and the l.h.s. is still small.

Function  $\frac{d \,L(\phi_z)}{d\,\phi_z}$ being small for $\phi_z \,<\,1$ has a singularity at 
$\phi\,\,\rightarrow\,\,1$ namely
$$
\frac{d \,L(\phi_z)}{d\,\phi_z}\,\,\,\longrightarrow\,\,\,\,\frac{1}{\left(\,1\,\,-\,\,\phi_z 
\,\right)^2}\,\,\,\,\,\,\,\,\,\mbox{for}\,\,\,\,\,\,\,\,\,\,\,\phi_z \longrightarrow\,\,\,1
$$
Therefore, the first requirement leads to the equation

\beq \label{EQSL3}
\frac{1}{\left( 1\,\,-\,\,\phi_z \,\right)^2}\,\,\,\frac{d^2\,\phi}{(d \hat{z})^2}\,\,=\,\,1.    
\eeq

For large $\z$ \eq{EQSL3} has a solution
\beq \label{EQSL4}
\phi(\hat{z})\,\,=\,\,\z\,\,-\,\,\ln \z
\eeq
which can be  verified  by explicit calculations. It should be stressed that $\phi(\z)$ of 
\eq{EQSL4} satisfies all conditions of a smooth function that has been used for the derivation of 
\eq{NLEPHI1}. \eq{SCDER} holds since $\phi_{zz}\, \,\approx\,\,1/z^2 \,\,\ll\,\,\phi^2_z 
\,\,\approx\,\, 1$. 
We can also check that the l.h.s. of \eq{NLEPHI1} is proportional to $1/z^2$ and it can 
be neglected.  

We check the validity of the solution of \eq{EQSL4} in more direct weay searching for the 
correction to \eq{EQSL4} due to a violation of \eq{SCDER}. The  need to do this, arises from 
the 
appearance of a contribution of the order of $1/z^2$ in $\phi^2_z$ which we cannot guarantee.
Searching for such corrections we replace \eq{SCDER} by a new equation
\beq \label{SCDER1}
\frac{d^n}{(d z)^n}\,e^{ - \phi(z)}\,\,=\,\, \left( ( - \phi_z)^n\,\,-\,\,n\,( - 
\phi_z)^{n-1}\,\phi_{zz} \right)\,e^{ - \phi(z)}
\eeq 

Using \eq{SCDER1} we obtain
\beq \label{COR1}
\frac{\chi(\gamma_{cr})}{1\,-\,\gamma_{cr}}\,\,\frac{d^2\,\phi}{(d \z)^2}
\,\,\,=\,\,\,\left(\,1\,\,-\,\,e^{- \phi(\z)}\,\right)\,\,-\,\,\frac{d \,L(\phi_z)}{d
\,\phi_z}\,\frac{d^2\,\phi}{(d \z)^2}\,\, +\,\,\frac{d^2 \,L(\phi_z)}{(d\,\phi_z)^2}
\,\,(\frac{d^2\,\phi}{(d \z)^2})^2
\eeq
instead of \eq{NLEPHI1}. \eq{EQSL3} has the following form
\beq \label{COR2}
\frac{1}{\left( 1\,\,-\,\,\phi_z \,\right)^2}\,\,\,\frac{d^2\,\phi}{(d \hat{z})^2}\,\,=\,\,1 + 
\frac{2}{\left( 1\,\,-\,\,\phi_{0,z} \,\right)^3}(\frac{d^2\,\phi_0}{(d \z)^2})^2
\eeq
where we denote by $\phi_0$ the solution of \eq{EQSL4}. Substituting $\phi_z = \phi_{0,z} 
\,+\,\Delta \phi_z$ in \eq{COR2} we obtain that
\beq \label{COR3}
\frac{d \Delta \phi_z}{d \z} \,\,=\,\,\frac{2}{\left( 1\,\,-\,\,\phi_{0,z} 
\,\right)}\,(\frac{d^2\,\phi_0}{(d \z)^2})^2\,\,\rightarrow\,\,\frac{2}{\z^3}
\eeq
One can see that \eq{COR3} leads to $ \Delta \phi_z\,\,\propto\,\,1/\z^2$ and the solution is 
\beq \label{COR4}
\phi(\z)\,\,=\,\,\z\,\,-\,\,ln\,\z \,\,+\,\,\frac{1}{\z}\,\,=\,\,\phi_0(\z) + O(1/\z)
\eeq

Hence,  \eq{EQSL4} is a solution.

It should be stressed that \eq{EQSL4} is a solution to the homogeneous equation ( \eq{NLEPHI1} with 
the l.h.s.  equal to zero). Therefore, the dependence on $Y$ in this solution stems only from the
matching of this solution to  the linear equation at $\z <0$. As has been shown (see 
\cite{GLR,MUT,IIM})
the solution of the linear equation behaves as 
\beq \label{NLEQSC}
N\left(\z \right)\,\,=\,\,N_0\,\,\exp \left( ( 1 - \gamma_{cr}) \z \right)
\eeq 
 We will 
discuss  this matching in more details later.

\subsection{The complete solution and matching with the pQCD domain}

Assuming that in \eq{NLEPHI1}, $\pz(\z)$ is a function of $\phi(\z)$ we can rewrite this equation in 
the form
\beq \label{EQPHI}
\L \,\frac{\chi(\gamma_{cr})}{ 1\,\,-\,\,\gamma_{cr}} \,\,+\,\,\frac{d L(\pz)}{d \pz} \R 
\,\pz(\z)\,\frac{d 
\pz(\phi)}{d\,\phi}\,\,=\,\,1\,\,-\,\,e^{ - \phi(\z)}
\eeq
 
Integrating \eq{EQPHI} with respect to $\phi$ we reduce this equation to the form which gives the 
implicit solution for $\z$ as a function of $\phi$.

$$
\frac{\chi(\gamma_{cr})}{1 - \gamma_{cr}}\,\h \L \pz \,-\,\pzn \R 
\,\,+\,\,\chi(1 - \pz)\,\pz 
\,-\,\chi(1 - \gamma_{cr})\,(1 - \gamma_{cr})\,-\,\int^{\pz}_{1 - 
\gamma_{cr}}\,\,\,\chi(\tilde{\pz})\,d\,\tilde{\pz}\,+\,\ln(\pz/\pzn)\,=
$$
\beq \label{EQPHI1}
=\,\,\,\phi + e^{- \phi} 
\,-\,\phi_0\,-\,e^{\phi_0}
\eeq
where $\pzn$ and $\phi_0$ are the initial conditions, namely, the value of function $\phi$ and 
$\pz$ 
at $\z = 0$. There are a number of  relations between them. The  first one comes from the matching of 
the 
logarithmic derivatives at $\z=0$:
\beq \label{INC1}
\frac{\h \L 1 - e^{- \phi_0} \R}{N_0}\,\,=\,\,1 \,-\,\gamma_{cr}
\eeq

The second relation  originates from the matching of the logarithmic derivatives for $d N/d z$
using the fact that the function $N$ has the following form near to the saturation line:$N(z)=N_0 
e^{(1 - \gamma_{cr})\,z}$ (see \eq{NLEQSC}), and that the geometrical scaling behaviour works even for 
negative $z$ 
\cite{IIM}. This relation is
\beq \label{INC2}
\pzn\,e^{ - \phi_0}\,\,=\,\,(1 \,-\,\gamma_{cr})\,\L 1 - e^{- \phi_0} \R 
\,\,.
\eeq

\begin{figure}[ht]
    \begin{center}
        \includegraphics[width=0.70\textwidth]{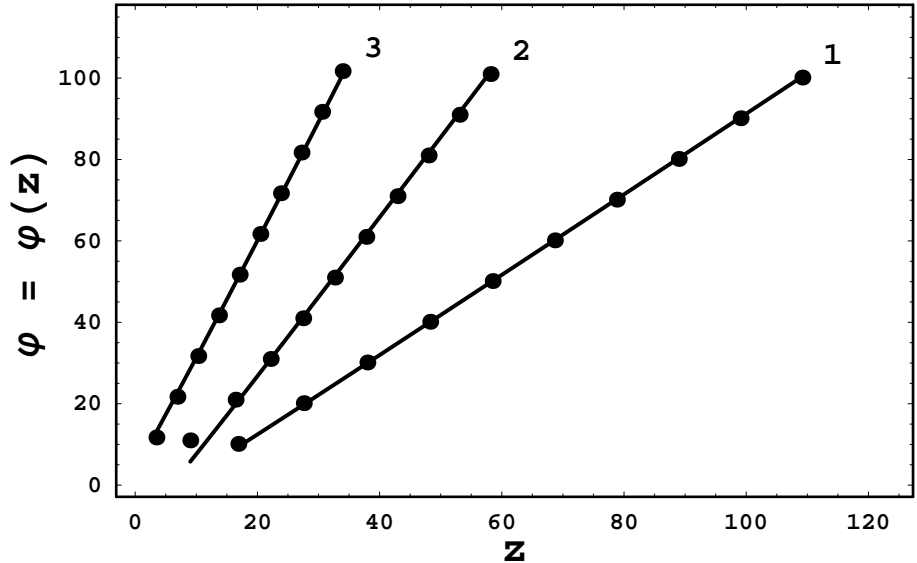}
        \caption{\it The numerical solution to \eq{EQPHI} as a function of the initial condition $N_0$.
Curves 1, 2 and 3 correspond to $N_0  = 0.1, 0.5, 0.65$, respectively. The lines are the asymptotic 
solution $\phi(z) = n\,z - \ln z + \ln \ln z$ with $n = 1,2,3$ for curves 1,2, and 3 , respectively.
}
\end{center}
\label{sol}
\end{figure}

Unfortunately,  with the initial conditions given by \eq{INC1} and \eq{INC2}, we could 
not solve \eq{EQPHI1}  analytically. \fig{sol} presents the numerical solution of this equation for 
positive 
$\z$ as a function of the initial  condition, namely, $N_0$ (see \eq{INC1} and \eq{INC2}).

One can see that the type of asymptotic behaviour depends on the value of $N_0$ at $z=0$. It has a 
simple explanation since $d L(\phi_z) d \phi_z \,\,\rightarrow ( n - \phi_z)^{-2}$. If $\phi_z$ at $z=0$ 
is smaller than unity we have the solution given by \eq{EQSL4}. However, if $2 > \phi_z > 1$ at $z=0$,
 we have $\phi_z(z) \,\rightarrow 2$ at large $z$. For $3 > \phi_z > 2$ the asymptotic behaviour is
 $ 3\,z  - \ln z$.  

It should be stressed that we cannot trust \eq{EQPHI1} for $z \,\rightarrow\,0$ since we cannot justify 
Eq.(2.10) in this region. In vicinity of $z=0$ this equation holds for the  function $N(z)$ rather than 
for  the 
function $\phi(z)$. Nevertheless if $N_0 \ll 1$ $\phi_0$ is small as well and our approach could be 
justified. From \fig{sol} one sees that at small $N_0$ we have the asymptotic behaviour given be 
\eq{EQSL4}. 

\section{Linearized equation in the saturation domain}

The key problem is the fact that the master equation is a  non-linear one. Therefore, generally 
speaking, we cannot conclude   that the 
 asymptotic behaviour of the solution can be written in the 
form (in space representation)
\beq \label{FSOL}
N(z)\,=\,1\, -\, e^{-\phi_1}\,-\,e^{ - \phi_2} 
\eeq
where $\phi_1$ is the solution of \eq{EQSL2},  while $\phi_2$ is given by \eq{EQSL4}.

However, the form of  \eq{PHIMR} shows us that the corrections are small in the region of large $z$
and ,  we can therefore  try to find the linear equation in this region for the function $e^{- \phi}$.
Denoting $e^{- \phi}$ by $S(r,Y;b)$ we obtain the following equation for $S$ from \eq{BK}:
\beq \label{BKS}
-\frac{\partial S\L r,Y;b \R}{\partial\,Y}\,\,=
\eeq
$$
\,\,\frac{C_F\,\as}{\pi^2}\,\,\int\,\frac{d^2 
r'\,r^2}{(\vec{r}\,-\,\vec{r}')^2\,r'^2}\, 
\L \,  S\L r,Y;\vec{b} \R \,\, -\,\, S\L r',Y;\vec{b} - \frac{1}{2}\,(\vec{r} - \vec{r}')
\R\, S\L
 \vec{r} -
\vec{r}',Y;\vec{b} - \frac{1}{2} \vec{r}'\R  \,
\R
$$

The first solution comes from the region of integration $r'\,\sim\,r \,\gg\,1/Q_s$ for  the second 
term on l.h.s. 
of the equation. In this region $S\,\ll\,1$,  and therefore, the non-linear term in \eq{BKS} can be 
neglected. The linear equation has a very simple form
\beq \label{LEQ1}
\frac{\partial S\L r,Y;b \R}{\partial\,Y}\,\,=\,\,-\,
\frac{C_F\,\as}{\pi}\,\,\int^{r^2}_{1/Q^2_s}\,\frac{d
r'^2\,r^2}{ |r^2 - r'^2|\,\,r'^2}\,
\,  S\L r,Y;\vec{b} \R \,\,=\,\,-\bas\,\,\ln(Q^2_s\,r^2)\,S\L r,Y;\vec{b} \R
\eeq
Substituting  a new variable $z$ (see \eq{Z})  \eq{LEQ1} has the  form:
\beq \label{LEQ2}
\frac{\chi(1 - \gamma_{cr})}{1 - \gamma_{cr}}\,\,\frac{d S(z)}{d z}
\,\,=\,\,- z\,S(z)
\eeq
and solution to  \eq{LEQ2} is very simple, namely,
\beq \label{LEQ3}
S(z)\,\,=\,\,\exp \left( - \frac{1 - \gamma_{cr}}{\chi(1 - \gamma_{cr})}\,\frac{z^2}{2} \right)
\eeq
The  simple derivation of this solution makes it transparent. 

The second solution comes from quite a different region in integration in \eq{BKS}, namely,
$r'$ or $r - r'$ are much smaller than $r$ and, basically, of the order of $1/Q_s$ since our 
solution depends on $z$. Let us assume for the sake of presentation that $  1/Q_s 
\,\approx\,|\vec{r} - \vec{r}'|\,\,
\ll\,\,r$. For such small distances we can replace $S(\vec{r} - \vec{r}',Y,b)$ by 1 ($S(\vec{r} - 
\vec{r}',Y,b) \,=\,1$). Indeed, in pQCD region $1 - S$ is the scattering amplitude and this 
amplitude is small.
Therefore, the linear equation that governs the asymptotic behaviour of $S$ has the following form:
\beq \label{LEQ40}
-\frac{\partial S\L r,Y \R}{\partial\,Y}\,\,=\,\,\frac{C_F\,\as}{\pi^2}\,\,\int\,\frac{d^2
r'\,r^2}{(\vec{r}\,-\,\vec{r}\;')^2\,r'^2}\,
\L \,  S\L r,Y \R \,\, -\,\,2\, 
  S\L
 \vec{r}',Y  \R_{|\vec{r} - \vec{r}\;'|\,\approx\,1/Q_s\,\ll\,r}  \,
\R
\eeq
In \eq{LEQ40} we neglected the impact paramer dependence of  $S$.

One can recognize in \eq{LEQ40} the BFKL equation, but with an  overall sign minus in front of the 
r.h.s.,
and with the restriction that the second term is valid only if $|\vec{r} - 
\vec{r}\;'|\,\approx\,1/Q_s\,\ll\,r$. The factor 2 in front of the second term on the r.h.s. of 
\eq{LEQ40},  
arise from the fact that we have the same contribution from the kinematic region where $r' 
\,\approx\,1/Q_s\,\ll\,r$.  We can rewrite \eq{LEQ40} in the form which is even closer to the BFKL 
equation by replacing the second term on r.h.s. of  \eq{LEQ40} by
\beq \label{MLEQ4}
\int\,\frac{d^2
r'\,r^2}{(\vec{r}\,-\,\vec{r}\;')^2\,r'^2}\,
  S\L
 \vec{r}',Y  \R_{|\vec{r} - \vec{r}\;'|\,\approx\,1/Q_s\,\ll\,r}  \,\,\rightarrow\,\,\,\int\,\frac{d^2
r'\,r^2}{(\vec{r}\,-\,\vec{r}\;')^2\,r'^2}\,\L\,
  S\L
 \vec{r}',Y  \R \,\,-\,\,\h S\L r,Y \R \,\R
\eeq
In \eq{MLEQ4} we  subtracted the region of integration $1/Q_s\,\ll\,
\,\,r' \,\ll\,\,r$ or/and $1/Q_s\,\ll\,
\,r' \,\gg\,\,r$. Indeed, rewriting the kernel in \eq{MLEQ4} for $r' <r$  in the form (after integrating 
over 
azimuthal angle)
\beq \label{MLEQ5}
\int\,\,\frac{d^2
r'\,r^2}{(\vec{r}\,-\,\vec{r}\;')^2\,r'^2}\,\,\rightarrow\, \pi \int^{r^2}\,d \,\,r'^2\,\L \frac{1}{r^2 
\,- \,r'^2}\,\,\,+\,\,\frac{1}{r'^2} \R
\eeq
one can see that the first term has been taken into account in \eq{LEQ40} since the region  $r' 
\,\approx\,r$ gives the dominant contribution to this term. In this statement we assume implicitly
that $S(r',Y)$ is steeply decreasing  in the saturation region. We have to  subtract the second term.
We can replace $r'$ by  $r$ in $S$ since this function is large only at $r' =   r$.  Returning  to the 
full kernel we obtain \eq{MLEQ5}. 

Taking \eq{MLEQ5} into account we can rewrite \eq{LEQ40} in the form
\beq \label{LEQ4}
-\frac{\partial S\L r,Y \R}{\partial\,Y}\,\,=\,\,\frac{C_F\,\as}{\pi^2}\,\,\int\,\frac{d^2
r'\,r^2}{(\vec{r}\,-\,\vec{r}\;')^2\,r'^2}\,
\L \,  S\L r,Y \R \,\, +\,\, \{ S\L r,Y \R\,\,-\,\,2\,\,
  S\L
 \vec{r}',Y  \R \}_{BFKL}  \,
\R
\eeq
The second term on the r.h.s. of \eq{LEQ4} is the kernel of the BFKL equation.

\eq{LEQ4} is very similar to the BFKL equation,  but it has a negative sign in front and also an 
additional term, proportional to $S\L r,Y \R$.

Using the form 
\beq \label{SZ}
 S(r',Y)\,\,=\,\,S(z)\,\,e^{ - \pz\,\ln(r'^2/r^2)}
\eeq 
we can see that \eq{LEQ4} can be reduced to \eq{NLEPHI1}. However, we do not need to take into account 
all the terms in \eq{NLEPHI1}. The main contribution comes from the first term on the r.h.s. of 
\eq{LEQ4} and 
from the second term in  the region of integration $r' \,>\,r$. Indeed, substituting \eq{SZ} in the 
second term on the r.h.s. of \eq{LEQ4}  the relevant contribution appears  as
\beq \label{SEQ1}
\int^{\infty}_{r^2}\,\,\frac{r^2\,\,d r'^2}{(r'^2 - r^2)\,r'^2}\,\L \,\L\frac{r'^2}{r^2}\R^{ - 
\pz}\,\,-\,\,1 \,\R\,\,=
\eeq
$$
\,\,\int^1_0\,\frac{d t}{1 - t}\,\L t^{\pz}\,\,-\,\,1 \R 
\,\,=\,\,\psi(1)\, -\, \psi(1 
\,+\,\pz)\,\,\rightarrow\,\frac{1}{1\,\, -\,\, \pz}
$$

Using the variable $z$ instead of $Y$ in \eq{LEQ4},  we obtain the simple equation which contains both 
discussed solutions:
\beq \label{SEQ2}
 \frac{\chi( 1 - \gamma_{cr})}{1  - \gamma_{cr}}\,\,\frac{d  S\L z \R}{d \,z}\,\,=\,\,\,z\, S\L z \R
\,\,+\,\,\frac{1}{1\,\, -\,\, \pz}\, S\L z \R
\eeq

Using $S\L z \R\,=\,e^{ - \phi(z)}$ we can easily reduce \eq{SEQ2} to an algebraic equation for $\pz$, 
namely,
\beq \label{SEQ3}
 \frac{\chi( \gamma_{cr})}{1  - \gamma_{cr}}\,\pz(z)\,\,=\,\,-\,z\,\,+\,\,\frac{1}{1\,\, -\,\, \pz}
\eeq
with the solution
\beq \label{SEQ4}
\pz^{(\pm)}\,\,=\,\,\frac{1}{2\,C}\,\L\,z\,+\,C\,\,\pm\,\,\sqrt{( 
z\,+\,C)^2\,\,+\,\,4\,C\,(1 - z)} \R
\eeq
where $C$ is given by \eq{C}.
At large $z$
\bea 
\pz^{(+)}\,\,&\rightarrow &\,\,\frac{z}{C}\,\,\label{SEQ5} 
\\
\pz^{(-)}\,\,&\rightarrow &\,\,1\,\,-\,\,\frac{1}{z} \,\label{SEQ6O
}
\eea
The second term on the r.h.s. of \eq{SEQ2} is very small if we substitute the solution  $
\pz^{(+)}$ of \eq{SEQ5} into \eq{SEQ2}. Therefore, this branch leads to
\beq \label{SS1}
S_1\L z \R \,\,\equiv\,e^{ - \phi_1}\,\,=e^{ - \int^z \,\,d\,z' \pz^{(+)}(z')}\,\,=\,\,\exp\L - 
\frac{z^2}{2\,C} \R
\eeq
The branch $ \pz^{(-)}$ leads a small l.h.s. of \eq{SEQ2} and therefore, the solution leads to  as a 
cancellation of the first and second terms on the r.h.s. of \eq{SEQ2}. 
The solution takes the form
\beq \label{SS2}
S_2\L z \R \,\,\equiv\,e^{ - \phi_2}\,\,=e^{ - \int^z \,\,d\,z' \pz^{(-)}(z')}\,\,=\,\,z\,\exp\L 
-
z\, \R
\eeq
The  solution $S_2(z)$ is the solution to the homogeneous equation ( see \eq{LEQ4} with
$\partial S(r,Y)/\partial Y = 0$ ). The entire dependence of this solution on energy ($Y$) stems from 
the matching with the solution of the linear equation  for negative $z$.  

\section{General solution for approaching the unitarity boundary }

Here, we are going to find a general solution to \eq{LEQ4}. We can solve this equation which is an 
equation of the 
BFKL -type, 
  using Mellin transform which we use for solution of the BFKL equation:
\beq \label{MLN}
S\L z \R\,\,=\,\,\frac{1}{2 \pi i}\,\int^{a + i \infty}_{a - i \infty} \,\,d\,f \,s(f)\,\,e^{f\,z}
\eeq
The main observation is that the function $(r'^2/r^2)^f$ is an eigenfunction of the BFKL kernel 
\cite{BFKL}. As we have seen (see \eq{SEQ1}) the integration in the region $r, > r$ leads to 
the eigenvalue
\beq \label{GS1}
\int^1_0\,\,\frac{d t}{1 - t}\,\L t^{-f}\,\,-\,\,1 \R\,\,=\,\,\psi(1) - \psi(1 -f)
\eeq
For $r' < r$ we have the following integral after integration over azimuthal angle :
\bea  
\int^{r^2}_0\,\,\frac{r^2\,\,d\,r'^2}{(r^2\,-\,r'^2)\,r'^2}\,\L \,\L\frac{r'^2}{r^2} 
\R^{f}\,\,-\,\,1 \R\,\,&=&\,\,-\, 
\int^{r^2}_{1/Q^2_s}\,\frac{d\,r'^2}{r'^2}\,\,+\,\,\int^{r^2}_{0}
\frac{\,\,d\,r'^2}{r^2\,-\,r'^2)}\,\L \L \frac{r'^2}{r^2} \R^{f - 
1}\,\,-\,\,1\,\R\,\,= \nonumber \\
\,\,-\,z\,\, +\,\,\int^1_0\,\frac{d\,t}{1 - t}\,\L t^{f - 1} 
-\,\,-\,\,1 
\R\,\,&=&\,\, -\,z\,\,+\,\,(\psi(1) - \,\psi(f))\,\label{GS2}
\eea

Collecting \eq{GS1} and \eq{GS2} we reduce \eq{LEQ4} to the form:
 \beq \label{EQSF}
C\,f\,s(f)\,\,=\,\,\frac{d s(f)}{d f}\,\,+\,\,\chi(f)\,\,s(f)
 \eeq 

To obtain  \eq{EQSF} we used the result  that a
multiplication by $z$ translates into operator $ - d/df$  for the Mellin transform.

\eq{EQSF} has the following solution
\beq \label{EQSFS1}
s(f)\,=\,A\,e^{X(f)}
\eeq
where
\beq \label{XS}
X(f)\,\,=\,\,\int^f_0\,d\,f' \,\L C\,f' - \chi(f')  \R\,\,=\,\,C\,\frac{f^2}{2} - 
2\,\psi(1)\,f\,\,  -\,\,\ln \Gamma(1 - f)\,\, +\,\, \ln \Gamma(f) 
\eeq
where constant $C$ and $\chi(f)$ are  defined in \eq{C} and \eq{CHI} , respectively, while   
 $\psi(1)$ is the Euler  constant, which is equal to 
0.577216.

The general solution to \eq{LEQ4}, which satisfies  the initial condition $S(z=0) = S_0$,
can be written in the form
\bea \label{GENSOL}
S\L z \R\,\,&=&\,\,S_0\,\int^{a + i\infty}_{a - i \infty}\,\frac{d f}{2 
\pi\,i}\,\,\frac{X'_f(f)}{X(f)}\,\exp\L X(f)\,\,+\,\,f\,z \R\,\, \\
 &=&\,\,S_0\,\int^{a + i\infty}_{a - i \infty}\,\frac{d f}{2\pi\,i}
\frac{C\,f - \chi(f) }{X(f)}\,\exp\L X(f)\,\,+\,\,f\,z \R \nonumber
\eea
where $a$ in the contour of integration is situated to the right of all singularities of the 
integrand \cite{BFKL}. 

At $z=0$ \eq{GENSOL} leads to
\beq \label{S0}
S\L z= 0  \R\,\,=\,\,S_0\,\frac{1}{2\,\pi\,i}\,\int\,\,\frac{d X}{X}\,e^{X}\,\,=\,\,S_0
\eeq

One can see two important  properties of the integrand in \eq{GENSOL}: (i) there are no 
singularities in it for $f>0$, since the singularities of $\Gamma(1 - f)$ in the exponent  
cancel  the 
singularities of $X'(f)$ in the numerator of \eq{GENSOL}; and (ii) 
contribution at the pole $f \,\rightarrow \,0$ vanishes since the numerator is equal to zero in 
this  point. This fact arises due to  the cancellation of single pole and double pole contributions. 
This 
 observation means that we correctly evaluated the  integral of \eq{GS2}. There is a  
problem in  that we can justify  the second term in the r.h.s. of \eq{LEQ4} only for 
$|\vec{r}\,-\,\vec{r}'|\,\,\approx\,\,1/Q_s\,\ll\,r$. However, at $f \,\rightarrow\,0$ the 
region of $|\vec{r}\,-\,\vec{r}'| \,\,\sim\,r$  also contributes since 
$\ln(r'^2/r^2)\,\approx\,1/f$. Strictly speaking we needed to subtract term $1/f$ from 
\eq{GS2}.  Our observation means that it was correct to keep  this term in 
the evaluation of the large $z$ behaviour of $S(z)$.

The integral in \eq{GENSOL} has two  sources for the asymptotic behaviour: the saddle point at large 
$f$ and the pole (singularities)  contributions at
$f\,\rightarrow\,-n$ where $n = 0,1,2 \dots$.

To calculate the saddle point contribution at large values of $f$ we replace all Gamma functions in 
\eq{GENSOL} by the asymptotic  expression. Therefore, the expression for $X(f)$ has the form
\beq \label{XLF}
 X(f)\,\,\,\rightarrow\,\,\,C\,\frac{f^2}{2}\, + 2f\,\ln(f) 
\,\,\,\,\,\,\,\,\,\,\,\,\,\,\,\,\mbox{at}\,\,f\,\gg\,1
\eeq
The equation for the saddle point is
\beq \label{SPXLF}
C\,f^{(+)}_{SP}\,\,+\,\,2\ln(f^{(+)}_{SP}) \,\,+\,\,z\,\,=\,\,0
\eeq
and
\beq \label{FSP+}
f^{(+)}_{SP}\,\,=\,\,- \frac{z}{C}\,\,-\,\,2\,ln(z)
\eeq
 
 Therefore, the saddle point contribution has the form
\beq \label{SSP+}
S^{+}\L z \R\,\,=\,\,S_0\,\sqrt{\frac{8\,\pi C}{z}}\,\exp\L - \h\,C\,(\frac{z}{C} \,+\,2\,\ln(z))^2 
\R
\eeq
which  is the same as our solution of \eq{SS1}, but obtained with  better accuracy. 

Near the singularity $ f = -n$ the integrand is
\beq \label{XX}
\frac{X'}{X}\,e^{X + f\,z}\,\,=\,\,\frac{(-1)^{n -1}}{(n-1)!}\frac{1}{\ln(n +f)}\frac{1}{(n + 
f)^2}\,e^{f\,z}\,\,=\,\,\frac{(-1)^{n -1}}{(n-1)!}\,\,\int^{\infty}_{0}\,d\,t\,\, \L n \,+\,f\,\R^{ - 2 + 
t}\,\,e^{f\,z}
\eeq

Closing the contour about the   point $j = - n$,  we see that the contribution is equal to
\bea \label{PON} 
\int^{\infty}_{0}\,d\,t \,\,\L n \,+\,f\,\R^{ - 2 + t}\,\,e^{f\,z}\,\,&=&\,\,e^{ - 
n\,z}\,\int^{\infty}_{0}\,d\,t\,\,\frac{1}{\Gamma(2 -t)}\,\,z^{1 - 
t}\,\,\nonumber \\
 &=&\,\,\,z\,\frac{1}{\ln (z)}\, \,e^{ -n\,z}
\eea
We used here the integral representation of the Euler gamma-function (see Ref. \cite{GR} {\bf 8.314}).

One can see from \eq{PON}, that only the first singularity is essential since all others are suppressed 
at 
large values of $z$.

However, they give more than the saddle point, and because of this, we have to sum all these 
contributions if we want to take into account the saddle point contribution as well.
The sum gives
\beq \label{SUM-}
S^{-}_{\Sigma}\L z \R\,\,=\,\,\frac{z}{\ln(z)}\,\L\,1 \,\,-\,\,\exp \L - \,e^{-z} \R \,\R
\eeq
Finally the behaviour of the solution  in the deep saturation region can be written as sum
\beq \label{FINS}
S_{asymp}\L z \R\,\,=\,\,S^{-}_{\Sigma}\L z \R \,\,+\,\,S^{+}\L z \R
\eeq
At large $z$ \eq{FINS} leads to
\beq \label{FINS2}
S_{asymp}\L z \R\,\,\,\rightarrow\,\,\,S_0\,\frac{z}{\ln(z)}\,\,e^{ - z}
\eeq

This solution in the form of \eq{FINS} has a very clear physical meaning. Each term is the 
suppression of gluon emission. Indeed, approaching the black disc limit, we  expect that an emitted gluon 
could not survive, except at the impact parameters close to the edge of the hadron disc \cite{LERY,KW,
IM}. Inside of the disc,  the dipole could elastically rescatter but cannot emit gluons. The $S^{+}$ 
term in \eq{FINS} describes the probability of dipole interactions without emission of a gluon in the 
leading twist. All other terms correspond to such a probability but for higher twists contributions in 
the BFKL equation, since the poles at $j = -n$ at $n >1$ correspond to the contribution of higher 
twists to the BFKL Pomeron \footnote{One of us (E.L.) thanks J. Bartels for very useful  discussions 
on higher 
twist contributions to the BFKL equation, that he had with him a number of  years ago.} (see for 
example 
Ref. \cite{DUR}). 

\section{Numerical check of the new solution}
 Despite the analytical calculations that led us to the solution in the form of \eq{FINS2}
 it is necessary to check that the homogeneous equation ( our master equation 
with zero l.h.s. ) has a solution. The main reason for such a check, is that the analytical
consideration is correct only for large $z \gg 1$,  and the matching between negative $z$ 
and large but positive $z$ is still out of  theoretical control. The second motivation 
for searching for  a numerical solution is the result of the numerical simulation by Salam 
\cite{MS} who  confirmed the solution of Ref. \cite{LT} rather  than the solution of 
\eq{FINS2}.

We search for the solution of the following equation
\begin{eqnarray}
  && \frac{\partial N(y,q)}{\partial y} =  \label{NUMeq1}  \\
        && \frac{\alpha_{S} \cdot N_{C}}{\pi} \; \left( \frac{1}{\pi} \int 
\frac{d^{2}q'}{(q - q')^{2}}
  \left[ N(y,q') \; - \; \frac{q^{2} }{q'^{\;2} \; + \; (q - q')^{2}} N(y,q) \right] \; - 
\; N^{2}(y,q)
 \right) \nonumber
\end{eqnarray}
This equation is the Balitsky-Kovchegov equation but in the momentum representation, where 
$q$ is a conjugated variable to $r$ while $Q$ is  a conjugated variable to $b$. For 
simplicity in \eq{NUMeq1}  we put $Q=0$.
After integration over angles \eq{NUMeq1}  reduces to the form
\begin{eqnarray}
         &&\frac{\partial N(y,q)}{\partial y} =  \label{NUMeq2}  \\
         &&\frac{\alpha_{S} \cdot N_{C}}{\pi} \; \left\{ \int_{0}^{\infty} d q'^{\;2}
 \left[\frac{N(y,q'^{\;2})}{| q'^{\; 2} - q^{2} |}  \; - \;
 \frac{q^{2}}{q'^{\;2}} \cdot \frac{N(y,q^{2})}{| q'^{\; 2} - q^{2} |} \; +
 \; \frac{q^{2}}{q'^{\;2}} \cdot  \frac{N(y,q^{2})}{\sqrt{4 q'^{\; 4} \; + \; q^{4}}}
  \right] \; - \; N^{2}(y,q^{2})  \right\} \nonumber 
\end{eqnarray}
We wish to solve this equation assuming the geometric scaling behaviour of the solution, 
namely, this solution is a function of the only one variable
\beq
\z\,\,\,=\,\,\ln\L Q^2_s(y,b)/k^2 \R \,\,=\,\,\,\bas \,\frac{\chi(\gamma_{cr})}{1
\,-\,\gamma_{cr}}\,\,(\,Y\,\,-\,\,Y_0\,)\,\,\,-\,\,\xi\,\,-\,\,\beta(b)\,\,;
\eeq
\noindent We can express \eq{NUMeq2} in terms of single variable $z$ using the following
expressions:
\begin{enumerate}
        \item $\;\;\; q^{2} \;  = \; e^{-z} \;\;\;$ and $\;\;\; dq^{2} \; = \; - \; e^{-z} 
\; dz \;\;\;
$
        \item $\;\;\; \frac{\partial}{\partial y} \; = \; \bar{\alpha_{S}} \; 
\frac{\chi(\gamma_{cr})}
{1 - \gamma_{cr}} \; \frac{\partial}{\partial z} $
\end{enumerate}
 Thus, finally we get following non-linear equation:
\begin{eqnarray}
         &&\frac{\partial N(z)}{\partial z} =   \nonumber \\
         &&\frac{1 - \gamma_{cr}}{\chi(\gamma_{cr})}  \; \left\{ \int_{\infty}^{\infty} d z'
\left[\frac{N(z') \cdot e^{-z'}}{| e^{-z'} - e^{-z} |}  \; - \;
 \frac{N(z) \cdot e^{-z}}{| e^{-z'} - e^{-z} |} \; + \;
\frac{N(z) \cdot e^{-z}}{\sqrt{(2 \; e^{-z'})^{\; 2} \; +
\; (e^{-z})^{2}}}   \right] \; - \; N^{2}(z)  \right\} 
\label{NUMeq3}
\end{eqnarray}

As we have mentioned  \eq{PHIMR}, we can find the scattering amplitude $N(z)$
 using  function $\phi(z)$.
We are going to demonstrate numerically, that  $\phi(z) = - \; (z - \log{z})$
(\eq{SS2}) minimizes the right hand side of \eq{NUMeq3}, i.e.
 it can be considered as solution for homogeneous equation:
\begin{eqnarray}
         0 \; = \; \frac{1 - \gamma_{cr}}{\chi(\gamma_{cr})}  \;
\left\{ \int_{-\infty}^{\infty} d z' \left[\frac{N(z') \cdot e^{-z'}}{| e^{-z'} - e^{-z} |}  \;
 - \; \frac{N(z) \cdot e^{-z}}{| e^{-z'} - e^{-z} |} \; +
\; \frac{N(z) \cdot e^{-z}}{\sqrt{(2 \; e^{-z'})^{\; 2} \; +
 \; (e^{-z})^{2}}}   \right] \; - \; N^{2}(z)  \right\}   \label{NUMeq4}
\end{eqnarray}

In order to perform required calculations we have to expand \eq{SS2} to the negative 
values 
of $z$.
 It is well known \cite{MUT,IIM,MP} that at the negative $z$, the scattering amplitude 
behaves as
 $N(z) = N_{0} e^{(1-\gamma_{cr})z}$. We also require continuation of $N(z)$ and $N'(z)$ 
at $z=0$.

We also require the matching of the two solutions at $z=0$ namely, the value of the amplitude 
$N(z)$ and its derivative  $N'(z)$  should be equal at $z=0$.

\beq
N(z)\,=\,\left\{
        \begin{array}{l}
\;\; N_{0} e^{(1-\gamma_{cr})z}  
\;\;\;\;\;\;\;\;\;\;\;\;\;\;\;\;\;\;\;\;\;\;\;\;\;\;\;\;\;\;\;\;\;\;\;\;\;
\;\;\;\;\;\;\;\;\;\;\;\;\;\;\;\;\;\;\;\;\;\;\;\;\;\;\;\;\;\;\;\;\;\;\;\;\;\;\;\;\;\;\;\;\;\;\;
\;\;\;\;\;
 z<0 \; ; \\  \\
\;\; N_{0} \; + \; \frac{1}{2} \; \int_{0}^{z} \left( 1 \; -
\; \beta \cdot \exp\left[- \; (z' \; - \; \log{[z' \; +
\; \frac{1 - 2 N_{0}(1-\gamma_{cr})}{\beta} ]} \; )\right] \right) dz'  \;\;\;\;\;\;\; z>0 
\; ;
        \end{array} \right.
\label{NUMeq5}
\eeq

\begin{figure}[ht]
    \begin{center}
\begin{tabular}{l l}
        \includegraphics[width=0.5\textwidth]{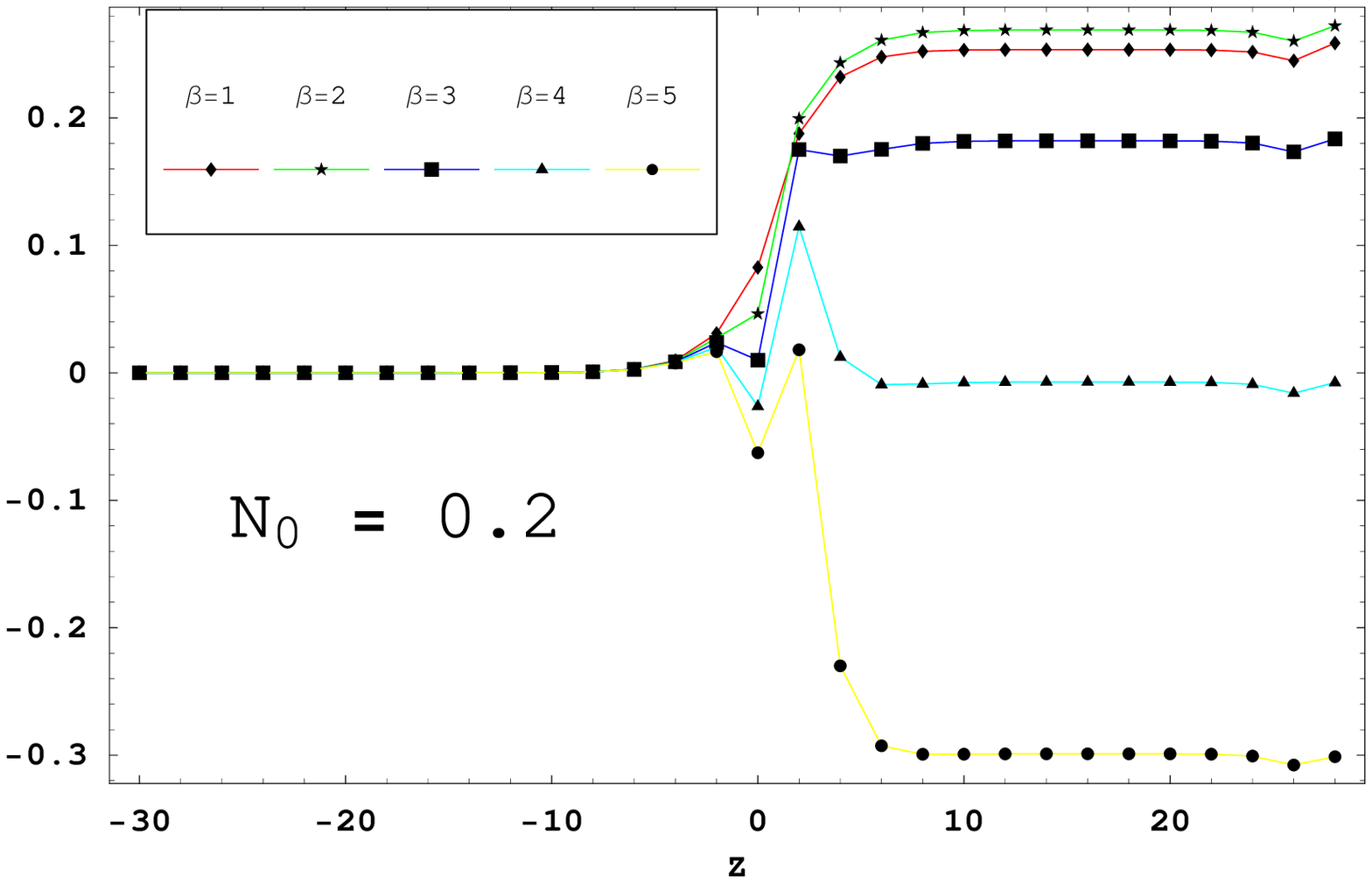} & 
\includegraphics[width=0.5\textwidth]{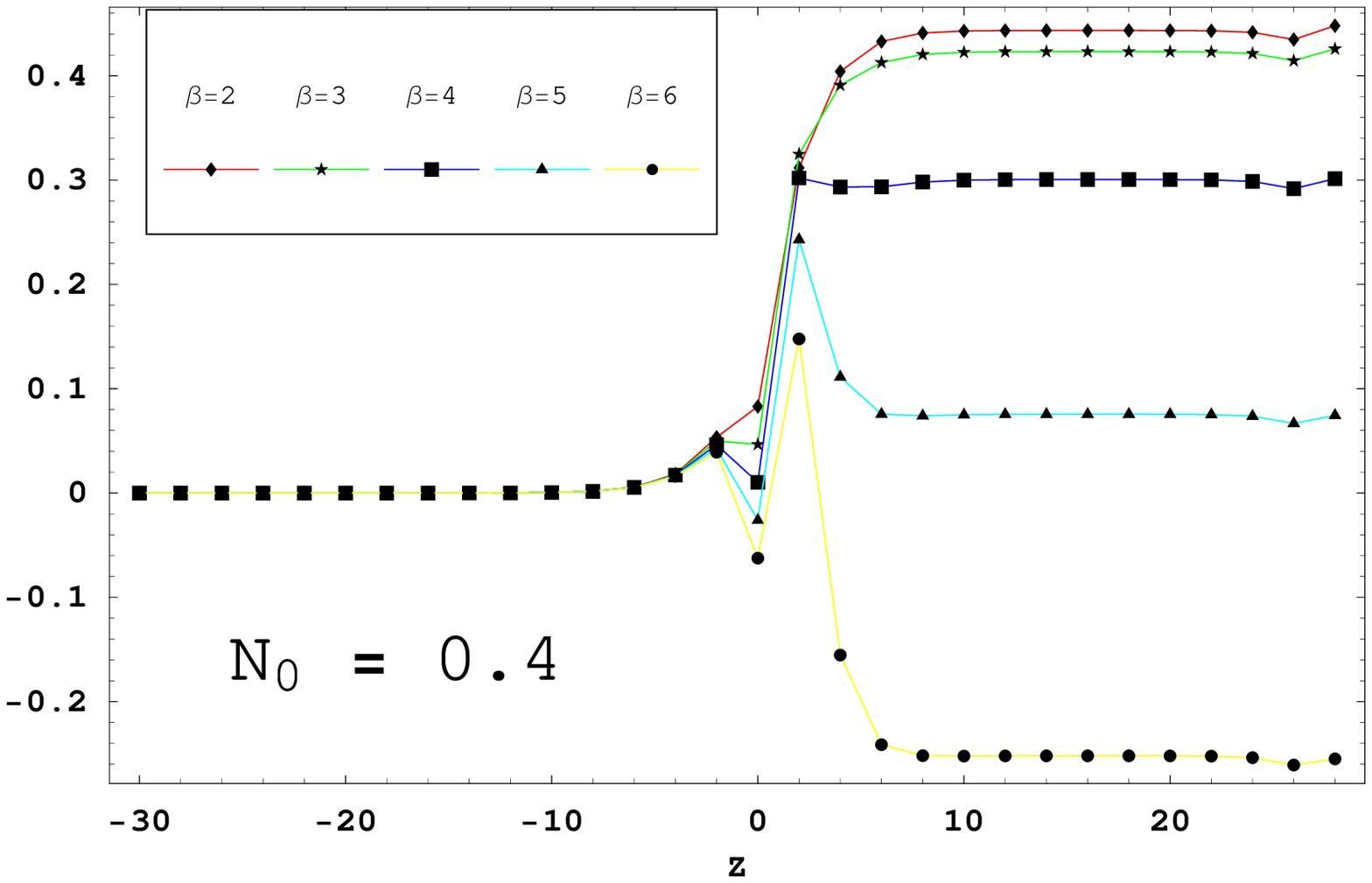}\\
             \fig{numsol}-a & \fig{numsol}-b\\
\end{tabular} 
\end{center}
       \caption{\it In  this plot we demonstrate results of numerical calculations
 of evaluation \eq{NUMeq4} for different values of $\beta$ with initial condition:
 $\; N(z=0)\;=\;0.2$ (\fig{numsol}-a) and  $\; N(z=0)\;=\;0.4$ (\fig{numsol}-b).}
\label{numsol}
\end{figure}

Parameter $\beta$, actually, accumulates all information which is beyond  the  precision 
of 
our solution.
As we can see from \fig{numsol}-a  variation of this parameter changes  
the value of the  r.h.s. of the  equation (see \fig{numsol})  that is 
calculated using \eq{NUMeq5}. One can see that the r.h.s. of the equation vanishes if we choose 
a value  of $\beta$. It means that $\phi = z - ln z$ really is a solution to the homogeneous 
equation which fulfills the correct boundary condition.

We hope that this numerical calculation will dissipate the lingering  doubts that the 
homogeneous equation has a solution, which   approaches   the asymptotic solution 
for high energy amplitudes.

\section{Beyond the Balitsky-Kovchegov equation}
In Ref. \cite{IIM} a new approach beyond the mean field approximation is proposed, based on the 
statistical interpretation of the high energy behaviour of the scattering amplitude. The main result 
of this paper can be understood  in the following way. 

The scattering amplitude for dipole-target interaction can be viewed as the average of the solution to 
the Balitsky-Kovchegov equation e with the saturation scale $Q_s$, over the saturation scale. 
The average saturation scale is given by the equation \cite{IMM}
\beq \label{QSIMM}
\frac{d \ln \bar{Q}^2_s}{d Y}\,\,=\,\,\bas \,\frac{\chi(\gamma_{cr}}{1 - \gamma_{cr}} 
\,-\,\frac{\pi^2\,\bas}{2}\,\frac{(1 - \gamma_{cr})\,\chi''(\gamma_{cr})}{\ln^2(1/\as^2)}
\eeq
and averaging should be performed with a Gaussian weight of variance 
\beq \label{SIGMA}
\sigma^2 \,= c\,\bas 
Y/\ln^3(1/\as) 
\eeq

Therefore the dipole amplitude is equal to \cite{IMM}
\beq \label{AIMM}
A(Y,r)\,\,=\,\,\frac{1}{\sqrt{2 \pi}\,\sigma}\,\int\,d \,\rho\,N\left(\bar{z},\rho \right)
e^{ - \frac{\rho^2}{2 \sigma^2}}
\eeq
where $ \bar{z}\,\,=\,\,\ln(r^2\,\bar{Q}^2_s)$ and $ \rho\,=\,\ln (Q^2_s/\bar{Q}^2_s)$.

This averaging drastically changes  the asymptotic behaviour of the first solution $N 
\,\propto e^{ - 
\frac{z^2}{2C}}$. It transforms into
\beq \label{FSIMM}
N \,\propto e^{ -
\frac{z^2}{2C}}\,\,\,\rightarrow\,A(\bar{z},Y) \,\propto\,\,e^{ - \,\frac{\bar{z}}{2\,\sigma^2}}
\eeq
Using \eq{SIGMA} one can see that \eq{FSIMM} leads to the following asymptotic behaviour at fixed $ 
r^2 \,<\,1/\bar{Q}^2_s$
\beq \label{FSIMM1}
A(\bar{z},Y)\,\,\,\propto\,\,\,\exp \left(- \frac{\bas}{2\,c} \,C^2 \,\ln^3 (1/as^3)\,\,Y \right)
\eeq

On the contrary  averaging does not dramatically  change \eq{FINS2} leading mostly to redefinition of 
the saturation scale, namely, the new scaling variable is equal to 
\beq \label{KSI}
\xi\,\,=\,\,\bar{z} \,\,-\,\,\h \,\sigma^2
\eeq
or, in other words, instead of saturation scale $\bar{Q}_s$ given by \eq{QSIMM} we have to introduce
a saturation scale
\beq \label{QSOUR}
\tilde{Q}^2_s\,\,=\,\,\bar{Q}^2_s\,e^{ -  \h \,\sigma^2} \,\,=\,\,\bar{Q}^2_s\,\exp \left( -  \h 
\,c\,\,\frac{\bas}{\ln^3(1/\as)}\,Y \right)
\eeq
In new variable the asymptotic behaviour of the scattering amplitude looks as
\beq \label{ASA}
A(Y,\xi)\,\,=\,\,\left( \xi  \,+\,\frac{3}{2}\,\sigma^2 \,\,-\,\,\frac{\sigma}{\sqrt{2 \pi}} \right)
e^{ - \xi}\,\,=\,\,\xi \,e^{ - \xi}
\eeq

Therefore, this solution shows the geometrical scaling behaviour while the first one displays no 
such scaling behaviour. 

\section{Summary}
We found that the Balitsky-Kovchegov  homogeneous equation (see \eq{BK} with the l.h.s. is  equal to 
zero) 
has a solution. This solution is determined by the singularities of the BFKL kernel in the anomalous 
dimension (Mellin conjugated variable to $\ln (r^2)$ where $r$ is the dipole size). This solution has 
the following form
\beq \label{SOLFIN}
N(r^2)\,\,=\,\,1 - C \exp \left( - z + \ln z\right)
\eeq
where $z \,\,=\,\,\ln (r^2/R^2)$ where $R$ is the arbitrary scale. $C$ is an arbitrary constant 
that we cannot calculate. Setting $r^2 = 1`/Q^2_S(Y)$ (
$Q_s$ is the saturation scale) on physical grounds we obtain that the solution to the 
Balitsky-Kovchegov
equation has the form:
\beq \label{SOLFIN1}
N(r^2)\,\,=\,\,1 -  Const\exp \left( - \ln(r^2 Q^2_s)\,\, + \,\,\ln  \ln(r^2 Q^2_s) \right)
\eeq
where $Const$ cannot be determined within our accuracy.

Since it depends on the singularities of the BFKL kernel it is very dangerous to make any 
simplification of this kernel. Indeed, in two approximate expressions  for  the BFKL kernel that 
have 
been 
suggested (see Refs. \cite{LT,MP}) the solution of \eq{SOLFIN1} is  missed. 
Formally speaking, this solution originates from the singularities of the BFKL kernel. In 
practical terms,  it means that we should be very careful  approximating  the full 
Balitsky-Kovchegov equation, by the simplified model of the BFKL kernel, namely, by diffusion
 approximation, in which this non-linear equation belongs to the 
Fisher-Kolmogorov-Petrovsky-Piscounov -type of equation (see Ref. \cite{MP}).   This issue needs 
further careful  investigation.


\section*{Acknowledgments:}
We want to thank Asher Gotsman and Al  Mueller  for very useful
discussions on the subject
of this paper. This research was supported in part  by the Israel Science Foundation,
founded by the Israeli Academy of Science and Humanities.



\end{document}